\documentclass[apj]{emulateapj}

\usepackage{amsmath}
\usepackage[normalem]{ulem}
\usepackage{color}

\newcommand{\nn}{\nonumber}
\newcommand{\be}{\begin{equation}}
\newcommand{\ee}{\end{equation}}
\def\lesssim{\buildrel < \over {_{\sim}}}
\def\gtrsim{\buildrel > \over {_{\sim}}}

\newcommand{\size}{1.0}  

\shorttitle{Collisionless shock in a partially ionized medium. 
II. Balmer emission}
\shortauthors{Morlino et al.}

\begin{document} 

\title{Collisionless shocks in a partially ionized medium. \\
II. Balmer emission}

\author{G. Morlino\altaffilmark{1}, 
        R. Bandiera\altaffilmark{1},  
        P. Blasi\altaffilmark{1,2}, 
        E. Amato\altaffilmark{1}}
\affil{$^1$INAF-Osservatorio Astrofisico di Arcetri, Largo E. Fermi,
                 5, 50125 Firenze, Italy}
\affil{$^2$INFN, Laboratori Nazionali del Gran Sasso, Assergi (AQ), Italy}

\begin{abstract}
Strong shocks propagating into a partially ionized medium are often associated
with optical Balmer lines. This emission is due to impact excitation of
neutral hydrogen by hot protons and electrons in the shocked gas.
The structure of such Balmer-dominated shocks has been computed in a previous
paper \cite[]{paperI}, where the distribution function of neutral particles 
was derived from the appropriate Boltzmann equation including coupling 
with ions and electrons through charge exchange and ionization. 
This calculation showed how the presence of neutrals can significantly modify the
shock structure through the formation of a \textit{neutral-induced} precursor
ahead of the shock. Here we follow up on our previous work and
investigate the properties of the resulting Balmer emission, with the aim of
using the observed radiation as a diagnostic tool for shock parameters.
Our main focus is on Supernova Remnant shocks, and we find that, for typical
parameters, the H$\alpha$ emission typically has a three-component spectral profile,
where: 1) a narrow component originates from upstream cold hydrogen atoms, 2) a
broad component comes from hydrogen atoms that have undergone charge exchange with
shocked protons downstream of the shock, and 3) an intermediate component is due
to hydrogen atoms that have undergone charge exchange with warm protons in the
neutral-induced precursor. The relative importance of these three components
depends on the shock velocity, on the original degree of ionization and on the
electron-ion temperature equilibration level. The intermediate component, which is
the main signature of the presence of a neutral-induced precursor, becomes
negligible for shock velocities $\lesssim 1500$ km/s. The width of the
intermediate line reflects the temperature in the precursor, while the
width of the narrow one is left unaltered by the precursor. In addition, we show
that the profiles of both the intermediate and broad components generally depart
from a thermal distribution, as a consequence of the non equilibrium
distribution of neutral hydrogen. 
Finally, we show that a significant amount of Balmer emission can be
produced in the precursor region if efficient electron heating takes place.

\end{abstract}

\keywords{ acceleration of particles -- atomic processes -- line:profiles --
ISM: supernova remnants }

\section{Introduction}
 \label{sec:intro}
It has been shown \cite[]{Chev-Ray78} that optical spectra dominated by
H$\alpha$ and other Balmer lines, as observed in some historical Supernova
Remnants (SNRs), may arise when an astrophysical shock propagates through a
partially ionized medium.

The Balmer emission, observed from these so-called Balmed-dominated shocks,
provides a powerful diagnostic tool to investigate the conditions existing
in the shock vicinity. The H$\alpha$ lines typically show two components,
resulting from excitation of neutral hydrogen due to the interaction with hot
protons and electrons in the shocked gas: a narrow-line component, whose width
is characteristic of the cold interstellar medium and that has been explained as
the result of direct excitation of neutral hydrogen atoms; and a much broader
component, associated to a second population of hydrogen atoms, created by
charge-exchange (CE) processes between the cold, still unshocked hydrogen and
the shocked protons. These hot atoms can be produced in an excited state or
can be excited by subsequent collisions with protons or electrons. Hence the
line width of the broad component traces the thermal velocity of shocked
protons and can be used to infer the shock velocity. Combining this estimate
with proper motion measurements, one can estimate the distance to the object.

Besides the shock speed, Balmer lines also represent a unique tool to
investigate the plasma physics of collisionless shocks. If both narrow and broad
components are detected, the relative intensity of the two lines can be
used to infer the ratio of electron-to-ion temperature just behind the shock,
providing information on the electron-ion equilibration mechanisms
\cite[]{Ghavamian07,vanAdelsberg08}.

One of the most intriguing aspects of Balmer emission is related to the
possibility of using the line shape and its spatial profile to check the
efficiency of SNR shocks in accelerating cosmic rays (CRs). If CR acceleration
is taking place in an efficient way, then  the widths of both the narrow and
broad line may be affected. In fact, when a sizable fraction of the ram pressure
is channeled into non-thermal particles, the plasma temperature behind the shock
is expected to be lower, and this should reflect in a narrower width of the
broad H$\alpha$ line. On the other hand efficient particle acceleration also
leads to the formation of a CR-induced precursor upstream, which heats the
ionized plasma before the shock. If the precursor is large enough, CE can occur
upstream leading to a broader narrow Balmer line. Remarkably, both signatures
seem to have been observed in Balmer-dominated shocks. 
For example \cite{Helder09} combined proper motion measurements of the
shock and broad H$\alpha$ line width for the remnant RCW 86 to demonstrate that
the temperature behind the shock is too low, thereby concluding that a sizable
fraction of the energy is being channelled into cosmic rays.
However, qualitatively similar features can also arise from different physical
processes. Therefore their observation can only be turned into reliable
information on the shock properties after a quantitative physical description of
the phenomenon is provided.

The basic theory of collisionless shocks in the presence of neutral particles
was first developed by \cite{Chev-Ray78} and \cite{Chevalier80} and further
refined by \cite{Smith91} and \cite{Ghavamian01}. These papers were however 
characterized by similar rather important limitations, mainly due to the
assumption that the distribution functions of both populations of neutrals are
Maxwellians. In fact, the main difficulty in describing the structure of a
collisionless shock propagating in a partially ionized medium is that neutrals
have no time to reach thermalization and cannot be treated as a fluid.
Steps forward in relaxing the fluid assumption have been made by
\cite{Heng07} and \cite{vanAdelsberg08} which considered the effect of multiple
CE events on the distribution function of hot neutrals. However these authors
limit their calculations to the region downstream of the shock, and consider
only the volume integrated distributions.

A proper description of the effects of the interactions between neutrals
and ions on the neutral distribution function requires a fully kinetic approach.
This approach has been implemented by \cite{paperI} (hereafter, Paper I)
where we derived simultaneously the neutral distribution function and the
hydrodynamic quantities for ions (which are, instead, treated as a fluid),
both upstream and downstream of the shock, by solving a Boltzmann equation 
including CE and ionization terms. The  main result of Paper I is that of providing a
mathematical and physical description of what we call the {\it neutral return
flux}: when fast, cold neutrals undergo CE interactions with the slower hot ions
downstream of the shock, some fraction of the resulting hot neutrals can cross
the shock and move upstream. The relative velocity between these hot neutrals
and the upstream ions triggers the onset of CE and ionization interactions that
lead to the heating and slowing down of the ionized component of the upstream
fluid. The system then tends to develop a shock precursor, in which the fluid
velocity gradually decreases from its value at upstream infinity, and even
more important, the temperature of ions in the upstream region increases as a
result of the energy and momentum deposition of returning neutrals.

The existence of a neutral return flux from downstream to upstream was
previously mentioned by \cite{Smith94} and \cite{Hester94} as a possible way to
explain the anomalous width of narrow Balmer lines observed in some SNRs
\cite[see e.g.][]{Sollerman03}, but no explicit calculation was carried out.

The width of such lines is in the 30-50 km/s range, implying a pre-shock
temperature around 25000-50000 K. If this were the ISM equilibrium temperature
there would be no atomic hydrogen, implying that the pre-shock hydrogen is
heated by some form of shock precursor in a region that is sufficiently thin so
that collisional ionization equilibrium cannot be established before the shock. 
Several mechanisms have been proposed to explain the broadening of the 
narrow line \cite[see][for a review]{Heng09}, but most of them can be ruled out on
theoretical grounds, leaving a CR precursor and/or a neutral precursor as the
most probable origin.

A first attempt at investigating the broadening of the narrow line component 
induced by the neutral precursor was made by \cite{Lim96}, using a simplified 
Boltzmann equation for neutrals in one dimension in both physical and velocity
space. The calculations were carried out for a shock speed of $225$ km/s, and
varying the initial ionization fraction from 0.5 to 0.99. The narrow line width
does not show any appreciable change due to the return flux.

In the present work we use the theory developed in Paper I and calculate the
profile of Balmer line emission in the presence of neutral return flux. In
particular we show that the neutral return flux is responsible for the emission
of a third intermediate line component, in addition to the narrow and broad
ones. This intermediate line is produced by hydrogen atoms that have undergone
CE with warm protons in the neutral precursor. Interestingly, there are
observations which suggest the existence of such intermediate component
\cite[]{Ghavamian00}, although the results might also be due to projection effects 
since the emission region is morphologically rather complex. At the present
time, spectral and spatial information are not sufficient to disentangle the
physical effect that we describe here from geometrical and projection effects
which could modify the line profiles.
We also show that the neutral-induced precursor is not able to broaden the width
of narrow H$\alpha$ line, confirming the first finding of \cite{Lim96}.

The paper is organized as follows: in \S\ref{sec:model} we summarize the kinetic
approach developed in Paper I and use it to describe the shock structure in the
presence of neutrals. We improve on previous work by both adding electrons
in the shock dynamics and including their contribution to the ionization of
neutrals. In \S\ref{sec:balmer} we write down the basic equations for the
calculation of the Balmer line emission and in \S\ref{sec:results} we illustrate
the main results in terms of spatial profiles of the total emission, of line
profile, as well as line intensity ratios. We also compare theoretical profiles
with profiles obtained by simulated observations, in order to derive the
observational requirements necessary in order to be able to detect the
intermediate component, and deviations from gaussianity in general. We conclude
in \S\ref{sec:conc}.

\section{Physical model}
\label{sec:model}

The basic model we consider consists of a plane-parallel shock with velocity
$V_{sh}$ that propagates into a partially ionized proton-electron plasma along
the $z$ direction, with a given fraction of neutral hydrogen at upstream
infinity. We neglect the presence of helium and other heavier chemical elements.

Protons and electrons are assumed to behave as fluids with temperatures
$T_i(z)$ and $T_e(z)$, respectively, with the same bulk velocity,
$v_i(z)=v_e(z)$, and the same number density, $n_i(z)=n_e(z)$. Their
distribution functions, $f_{i}(\vec v,z)$ and $f_{e}(\vec v,z)$, are assumed to
be Maxwellian at each position $z$.
Neutral hydrogen interacts with protons and electrons through CE
and ionization. The hydrogen distribution function, $f_{N}(\vec v,z)$, can be 
described using the Boltzmann equation
\begin{equation}
\frac{\partial f_{N}}{\partial t} + \vec v \cdot \nabla f_{N} = \beta_{N} f_{i} 
 - \beta_{i} f_{N} - \beta_e f_{N},
\label{eq:vlasov}
\end{equation}
where the collision terms $\beta_k f_l$ represent the interaction (due to CE and
ionization) between the species $k$ and $l$. The interaction rate $\beta_k$ is
formally written as
\begin{equation} \label{eq:beta_k}
\beta_k (\vec v,z) = \int d^{3} w \, v_{rel} \, \sigma(\vec v_{rel})
                  f_{k}(\vec w,z) \,,
\end{equation}
where $v_{rel} = |\vec v- \vec w|$ and $\sigma$ is the cross section for the
interaction process. More precisely, $\beta_N$ is the rate of CE of an ion that
becomes a neutral, $\beta_i$ is the rate of CE plus ionization of a neutral due
to collisions with protons, while $\beta_e$ is the ionization rate of neutrals
due to collisions with electrons. Eq.~(\ref{eq:vlasov}) is used to calculate
$f_N$ starting from the distribution of charged species (protons and electrons),
under the assumption of stationarity ($\partial f_N/\partial t=0$).

The dynamics of protons and electrons coupled with neutrals can be described 
very generally through conservation equations of mass, momentum and energy:
\begin{equation} \label{eq:rh1}
 \frac{\partial}{\partial z} \left[ (\rho_i + \rho_e) v_{i} + F_{mass}
  \right]=0 \,,
\end{equation}
\begin{equation} \label{eq:rh2} 
 \frac{\partial}{\partial z} \left[ (\rho_i + \rho_e) v_{i}^{2} + P_{i} + P_{e}
 + F_{mom} \right]=0 \,,
\end{equation}
\begin{equation} \label{eq:rh3}
 \frac{\partial}{\partial z} \left[ \frac{1}{2} (\rho_i + \rho_e) v_{i}^{3} + 
  \frac{\gamma_{g} (P_{i}+P_{e}) v_{i}}{\gamma_{g}-1}+ F_{en} \right]=0 \,,
\end{equation}
where $F_{mass} = m_H \int d^{3} v v_{\parallel} f_{N}$, $F_{mom} = m_H \int
d^{3} v v_{\parallel}^{2} f_{N}$ and $F_{en} = m_H/2 \int d^{3} v v_{\parallel}
(v_{\parallel}^{2}+v_{\perp}^{2}) f_{N}$ are respectively the fluxes of mass,
momentum and energy of neutrals along the $z$ direction. 
Usually the dynamical role of electrons is neglected due to their small mass.
However, collective plasma processes could contribute to equilibrate electron
and proton temperatures to some level. If this equilibration is very efficient,
then the electron pressure can no longer be neglected and the total gas pressure
needs to include both proton and electron contributions, namely $P_{g}= P_{i} +
P_{e} = P_{i}(1+\beta)$, where $\beta(z)\equiv T_e/T_i$ is the electron to
proton temperature ratio and is taken as a free parameter.

The solution of Eqs.(\ref{eq:vlasov})-(\ref{eq:rh3}) is described in detail in
Paper I. The only difference here is the presence of electrons, which has been
previously neglected because it does not affect the shock dynamics, unless 
$T_e\sim T_i$. The importance of introducing the electron contribution in this
context comes from the fact that Balmer emission is very sensitive to both $T_e$
and $T_i$ \cite[]{Heng07,vanAdelsberg08}. Therefore, even very partial
equilibration can produce observational consequences in the line emission, as we
show in \S\ref{sec:results}.

The level of electron-ion equilibration is one of the open questions in
collisionless shock physics and H$\alpha$ observations present us with a unique
tool to investigate this aspect. Theoretically the equilibration process
is far from being understood and anything between total absence of interaction
and full equilibration has been proposed \cite[see][for a review]{Rakowsky05}.
Moreover the electron-ion equilibration level is expected to change between
upstream and downstream of the shock, because the plasma conditions in these two
regions are totally different in terms of temperature and turbulence properties.
As a consequence, we distinguish between upstream and downstream using two
separate parameters, $\beta_{up}$ and $\beta_{down}$, respectively. 

We stress that a crucial assumption of our calculation is that new protons 
produced at position $z$ by CE and ionization instantly thermalize
with other protons. As discussed in Paper I, such an assumption, which is
especially important in the upstream region, is rather delicate: protons might
isotropize and yet not thermalize with the rest of protons in the plasma. A
dedicated effort using numerical particle simulations should be used to address
this important issue.

In principle, the dynamics of electrons in the upstream region could also be
affected by electrons ejected due to the ionization of neutrals that have returned
upstream. This effect can however be neglected for the calculation of Balmer lines: 
in Paper I we showed that the fraction of returning neutrals is $\lesssim 5\%$.
Let us assume that all these neutrals are ionized in the upstream region. To a
good approximation, stripped electrons can be expected to have the same velocity
of their parent atoms, which, in turn, have a mean velocity of the order of the
downstream proton thermal speed. Hence the temperature of stripped electrons is
$T_{e,in} \simeq 3/8 V_{sh}^2 m_e/k_B \approx 6.2\times 10^5 (V_{sh}/5000 \rm
km/s)^2$K. Now, assuming that the newly generated electrons reach equilibrium
with preexisting electrons, the final temperature is of the order of $(n_0 T_0
+n_{in} T_{in})/(n_0+n_{in}) \approx 2.5\times 10^4$K, where we use $V_{sh}=
5000$ km/s and an initial ionization fraction of 50\%. This temperature is
too low to be relevant for Balmer emission. In fact, electrons can contribute
only if their velocity is $\gtrsim 2000$ km/s, namely if $T_e \gtrsim 10^5$K.

\section{Calculation of Balmer line emission}
\label{sec:balmer}
Once the distribution functions of hydrogen, protons and electrons are known at
each position, the calculation of line emission is quite straightforward,
provided that the relevant cross sections are known. Here we concentrate
on the Balmer H$\alpha$ line, which results from the hydrogen deexcitation from
level $3s$ and $3d$ to $2p$ as well as from $3p$ to $2s$. The latter case is
complicated by the fact that the $3p$ level can also decay into $1s$, producing
Ly$\beta$ photons. Depending on the optical depth of the medium, these photons
can either escape the system or be reabsorbed by ground-state hydrogen and
eventually reemitted as H$\alpha$ photons. In the literature the optically thin
and optically thick cases are usually labelled as Case A and Case B,
respectively, and the total H$\alpha$ production rate is written as:
\begin{equation} \label{eq:Ralpha}
 R_{H\alpha} = R_{H(3s)} + R_{H(3d)} + B_{3p,2s} R_{H(3p)} \,,
\end{equation}
where $R_{H(3l)}$ is the production rate of hydrogen excited at level $3l$
and the factor $B_{3p,2s}$ is the fraction of transitions from $3p$ to $2s$,
which is $\approx 0.12$ in the optically thin case (Case A) while it becomes
unity in the optically thick case (Case B) \cite[]{vanAdelsberg08}.

The conversion efficiency from Ly$\beta$ to $H\alpha$ photons depends on the
shock speed, the electron-ion temperature ratio and the pre-shock ionization
fraction. It was first computed by \cite{Chevalier80}, who found that in
conditions appropriate for many Balmer dominated shocks the emission from cold
hydrogen is generally optically thick while the emission from hot hydrogen is
close to be optically thin. In this work, following that result, we adopt
$B_{3p,2s}=0.12$ (1) for the emission produced by hot (cold)
hydrogen. 

In order to calculate the H$\alpha$ emission for both Case A and Case B, we need
to compute the different production rates of hydrogen excited to the sublevels
$3s$, $3p$ and $3d$. The excited hydrogen is mainly produced by two different
processes: collisions with protons and electrons, and CE reactions
leaving the hydrogen atom in an excited state. For the sake of clarity here we
neglect further contributions due to collisions with helium. The production rate
of $H(3l)$ at a fixed position $z$ reads:
\begin{eqnarray} \label{eq:H(3l)}
 R_{H(3l)}(\vec v,z) = \int d^3 w \, v_{rel} \, f_{H}(\vec v,z) \times \nn \\
      \left[ f_i(\vec w,z)  \, \sigma_{ex(i)}^{tot(3l)}(v_{rel}) 
      + f_e(\vec w,z)  \, \sigma_{ex(e)}^{tot(3l)}(v_{rel}) \right]  \nn \\
   + \int d^3 w \, v_{rel} \,
      f_i(\vec v,z) f_{H}(\vec w,z) \, \sigma_{ce}^{tot(3l)}(v_{rel}) \,.
\end{eqnarray}
Notice that the hydrogen distribution function $f_{H}$ includes only atoms at
the ground level $1s$. In fact, collisional deexcitation can be neglected
because it occurs on a typical time scale $\tau_{coll} \simeq (n_i \sigma_{coll}
V_{sh})^{-1} \simeq 10^7 {\rm sec}$ (for $n_i= 1\, {\rm cm^{-3}}$ and
$V_{sh}=1000\, {\rm km/s}$) which is much longer than spontaneous deexcitation,
whose time scale (from state $n$ to state $m$) is $\tau = \hbar/\Delta E= 4.8
\times 10^{-15} (n^{-2}-m^{-2})\, {\rm sec}$. Hence atoms decay to the ground
state before undergoing any further collision.

In Eq.~(\ref{eq:H(3l)}) we use the total excitation and CE cross sections for
the atomic sublevel $3l$, $\sigma_{ex}^{tot(3l)}$ and $\sigma_{ce}^{tot(3l)}$,
which take into account the direct excitation processes, $1s \rightarrow 3l$,
plus contributions coming from atoms excited to higher levels with $n>3$ that
subsequently decay to the state $3l$. Formally this total cross section can be
written in the following form:
\begin{equation} \label{eq:sigma}
 \sigma^{tot(3l)} = \sum_{n',l'}^{\infty} \, B_{n'l',3l} \,
 \sigma_{1s \rightarrow n'l'} \,.
\end{equation}
where $\sigma_{1s\rightarrow n'l'}$ is the cross section for excitation from
the ground state to the level $n'l'$ and the $B_{n'l',nl}$ are the cascade
matrix elements representing the probability that a hydrogen atom excited to state
$n'l'$ will make a transition to state $nl$ (with $n<n'$) via all cascade
routes. 

We treat CE, excitation and ionization between electrons, protons and hydrogen
atoms using the cross sections of \cite{Barnett90}, \cite{Belkic92},
\cite{Janev93}, \cite{Balanca98} and \cite{Harel98}\footnote{Many of these cross
sections can be found in the International Atomic Energy Agency website:
\texttt{http://www-amdis.iaea.org/ALADDIN/} }. For some of these cross sections
we adopt the fitting functions provided by \cite{Heng-Sun08} and
\cite{Tseliakhovich12}.
At the time of writing, the CE cross sections for sublevels with different
angular momentum states are known only up to the level $n=4$. We estimate
that neglecting the contribution of higher levels entails an error around $5\%$
in the calculation of the total $\sigma_{ce}$, therefore in the following we
ignore levels with $n>4$. Hence, for the CE process, the cross sections reduce
to the following:
\begin{align} 
 \sigma_{ce}^{tot(3s)} &\simeq \sigma_{ce,1s \rightarrow 3s} 
   + B_{4p,3s} \, \sigma_{ce,1s \rightarrow 4p} \,, \label{eq:sigma_ce1}   \\
 \sigma_{ce}^{tot(3d)} &\simeq \sigma_{ce,1s \rightarrow 3d} 
   + B_{4d,3d} \, \sigma_{ce,1s \rightarrow 4d} \,, \label{eq:sigma_ce2}  \\
 \sigma_{ce}^{tot(3p)} &\simeq \sigma_{ce,1s \rightarrow 3p} 
   + B_{4s,3p} \, \sigma_{ce,1s \rightarrow 4s} \nn \\
   &\quad+ B_{4d,3p} \, \sigma_{ce,1s \rightarrow 4d}\,,  \label{eq:sigma_ce3}
\end{align}
where we use the values of $B_{4l',3l}$ as listed by \cite{Heng-Sun08} (see
their Table 3). An identical approach is adopted to calculate the impact
excitation by protons, restricted to levels 3 and 4. We adopt the impact
excitation cross sections calculated by \cite{Balanca98} and
\cite{Tseliakhovich12} for the sublevels $3l$ and $4l$ respectively.
Unfortunately these works provide cross sections only for a limited range of
impact energies, 1--100 keV and 5--80 keV respectively, which means that the 
relative speed between protons and hydrogen atoms can be respectively in the
range $[v_1,v_2] = [438,4377]$~km/s and $[978,3914]$~km/s. Outside these
velocity ranges we estimate the sublevel cross sections, $\sigma_{ex(i)}^{nl}$,
using the total cross sections, $\sigma_{ex(i)}^{n}$ as taken from
\cite{Janev93}, in the following way:
\begin{equation}
 \sigma_{ex(i)}^{nl}(v>v_2) = \sigma_{ex(i)}^n(v) 
   \frac{\sigma_{ex(i)}^{nl}(v_2)}{\sigma_{ex(i)}^{n}(v_2)} 
   \left(1 + \epsilon_n \right) \,,
\end{equation}
where the coefficients $\epsilon_n$ are chosen
in such a way as to have $\sum_{l} \sigma_{nl}(v_{1(2)}) = \sigma_n(v_{1(2)})$, and
their values are of the order of few percent.
A similar approximation has been used also for $v<v_1$. 

Also the impact excitation by electrons is limited to sublevels 3$l$ and 4$l$
and expressions similar to (\ref{eq:sigma_ce1})-(\ref{eq:sigma_ce3})
hold for $\sigma_{ex(e)}^{tot(3l)}$ as well. In this case we use the cross
sections provided by \cite{Bray95}, computed using the convergent close-coupling
method. As for the previous case, the error in the total cross section produced
by excluding higher excited levels is around 5\%.

In order to compute the spatial emissivity profile of the H$\alpha$ emission we
need to integrate Eq.~(\ref{eq:Ralpha}) in the velocity directions orthogonal to
the line of sight. From the observational point of view, most cases refer to 
shocks viewed edgewise because of limb brightening. In such cases, assuming a
pure plane shock (i.e. neglecting curvature effects), the edge-wise emissivity
profile results from the following integration:
\begin{eqnarray} \label{eq:Halpha-em2}
 F_{H\alpha}(z,v_x)
 = \iint dv_z dv_y \left[ R_{H(3s)}(z,\vec v)\right.\qquad\qquad\nn \\
     \left.  + R_{H(3d)}(z,\vec v)+ B_{3p,2s} R_{H(3p)}(z,\vec v) \right]  \nn \\
 = 2 \int_{-\infty}^\infty dv_z
    \int_{v_x}^{\infty} \frac{dv_\bot}{\sqrt{1-(v_x/v_\bot)^2}}\quad\;\;  \nn \\ 
     \left[ R_{H(3s)}(z,v_z,v_\bot)+ R_{H(3d)}(z,v_z,v_\bot)\right.   \nn \\ 
\left. + B_{3p,2s} R_{H(3p)}(z,v_z,v_\bot)\right] \,,
\end{eqnarray}
where $z$ is the direction of shock propagation, $x$ is the direction along the
line of sight and $y$ is orthogonal to the $(x,z)$ plane.
In the second equality we use $v_\bot \equiv ({v_x^2+v_y^2})^{1/2}$. 
Starting from Eq.~(\ref{eq:Halpha-em2}) we can obtain more useful integrated
quantities, which can be directly compared with observations, namely the spatial
emissivity profile, $\xi_{H\alpha}(z)$, the volume-integrated line profile,
$\phi_{H\alpha}(v_x)$, and the total line strength, $I_{H\alpha}$, respectively defined as
\begin{eqnarray} \label{eq:z-profile}
 \xi_{H\alpha}(z) = \int_{-\infty}^{\infty} dv_x F_{H\alpha}(z,v_x) \,, \nn \\
 \phi_{H\alpha}(v_x) = \int_{-\infty}^{\infty} dz\, F_{H\alpha}(z,v_x) \,, 
 \;  {\rm and} \nn \\
 I_{H\alpha} = \iint dv_x dz \, F_{H\alpha}(z,v_x) \,.
\end{eqnarray}
In the next section we will investigate how these observables can be used to 
test the presence of a neutral-induced precursor and to infer the ambient
parameters (as the ionization fraction, the shock speed and the electron-proton
equilibration level).

\section{Results}
\label{sec:results}
In this section we illustrate the main results of our kinetic calculation
concerning the H$\alpha$ emission. It is well known that Balmer emission is
highly sensitive to plasma density, shock velocity, initial degree of ionization
and electron-ion equilibration level. Our aim is to illustrate how to disentangle
different effects produced by these quantities.
The main observable quantities are the spatial profile, the line profile and the
relative intensity of the broad and narrow lines. Below we discuss the
interpretation of these quantities and their relation with physical parameters,
comparing our results with previous work.

For all cases discussed, we fix the temperature at upstream infinity to $10^4$K,
because larger values are incompatible with the presence of neutral hydrogen,
while for lower values the results do not change significantly. 
Also the total upstream numerical density is fixed to $n_{tot}= 0.1 {\rm
cm^{-3}}$, while the plasma ionization fraction is assumed to be 50\%, unless
otherwise specified. We notice that the total density can be factorized out
in Eqs.(\ref{eq:vlasov})-(\ref{eq:rh3}), hence a change in the total density
only reflects in a change of the length scales of the problem, while all other
quantities remain unchanged. 

As we already pointed out, the electron-to-proton equilibration level, $\beta$, is
expected to change between upstream and downstream of the shock, hence we use
two independent parameters, $\beta_{up}$ and $\beta_{down}$, respectively. We
will focus mainly on the two extreme cases of full equilibration (FE), where
protons and electrons share the same temperature everywhere (i.e. $\beta_{up}=
\beta_{down}=1$) and no equilibration at all (NE), which corresponds to the situation
where electrons and protons do not interact at all. In the
latter case the electron temperature is equal to $10^4$K in the entire upstream
region and $\beta_{down}=m_e/m_p$ in the downstream. Intermediate equilibration
cases will be also discussed.

\subsection{Spatial emission} \label{sec:spatial}
\begin{figure}
\begin{center}
\includegraphics[width=\size\linewidth]{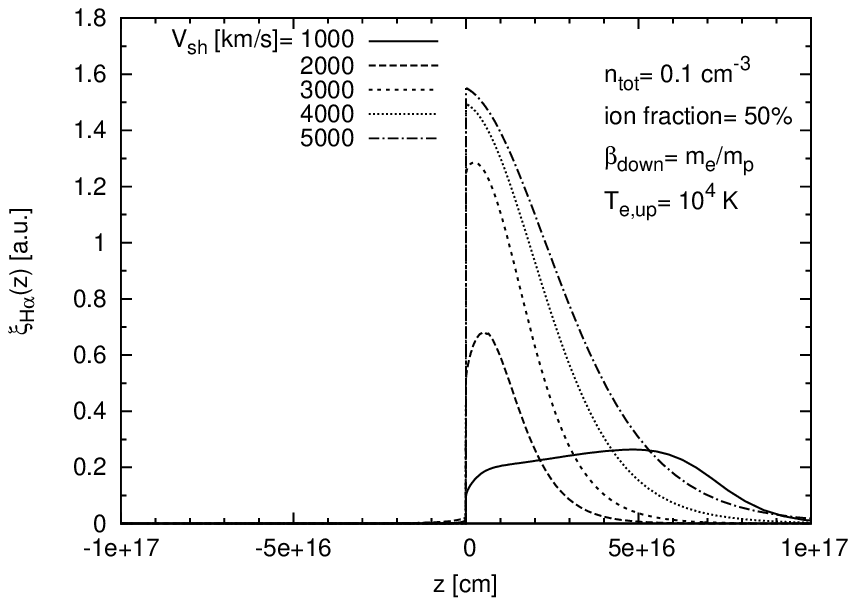}
\includegraphics[width=\size\linewidth]{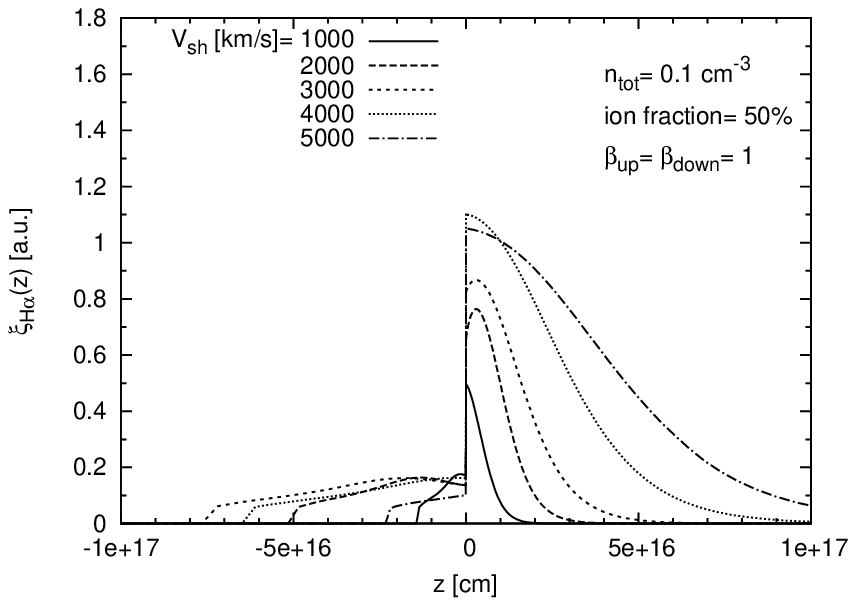}
\end{center}
\caption{Spatial emissivity profile of the H$\alpha$ line for $n_{tot}=0.1 {\rm
cm^{-3}}$, 50\% ionization fraction and for different values of shock velocity,
as specified in each panel.
Two extreme cases of electron-ion temperature equilibration are shown:
NE ({\it upper panel}) and FE ({\it lower panel}), i.e.
$\beta_{up}=\beta_{down}=1$.}
\label{fig:Halpha_z}
\end{figure}

\begin{figure}
\begin{center}
\includegraphics[width=\size\linewidth]{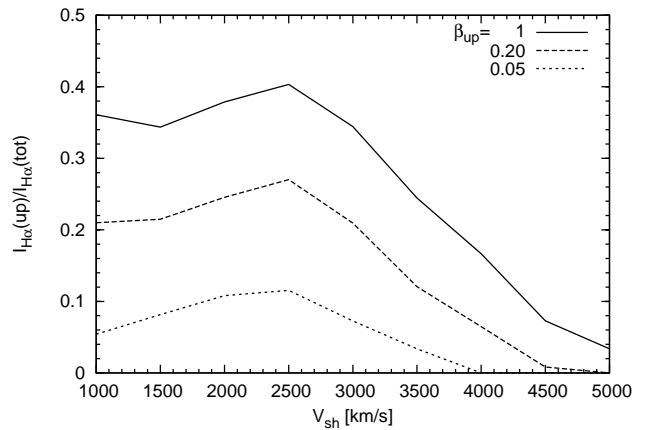}
\end{center}
\caption{Fraction of the total emissivity coming from the upstream as a function
of shock velocity and for three different assumptions on the degree of electron-proton
equilibration upstream ($\beta_{down}$ is fixed to 1 for all cases).}
\label{fig:Halpha_up}
\end{figure}

\begin{figure}
\begin{center}
\includegraphics[width=\size\linewidth]{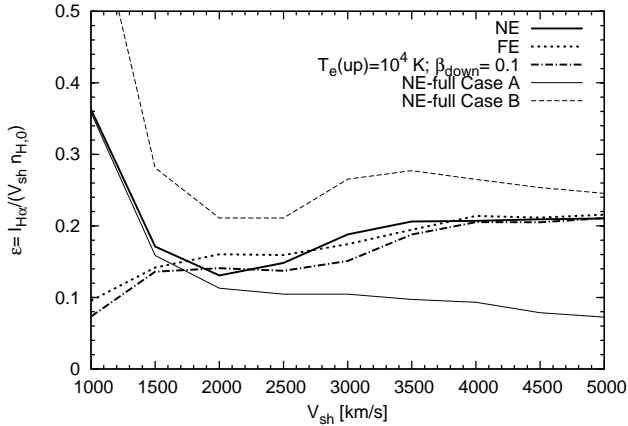}
\end{center}
\caption{Efficiency of the H$\alpha$ emission (i.e. number of photons
emitted for each hydrogen atom crossing the shock) as a function of the shock
velocity and for different degrees of electron-ion equilibration efficiency. The
thick lines refer to the cases of non-equilibrium (solid), full equilibrium (dotted)
and partial equilibrium with $\beta_{down}=0.1$ (dot-dashed). For these cases the plasma is
assumed to be optically thin for Ly$\beta$ photons produced by hot hydrogen
atoms, but optically thick for photons produced by cold hydrogen atoms. The thin
lines, instead, show NE cases assuming a medium that is either optically thin (Case A) or
optically thick (Case B) to Ly$\beta$ photons.}
\label{fig:Halpha_eff}
\end{figure}

Fig.~\ref{fig:Halpha_z} shows the spatial emissivity profile,
$\xi_{H\alpha}(z)$, for different shock velocities and for the two extreme
assumptions for the electron-ion equilibration efficiency: the upper panel
shows the NE case, while the lower panel shows the FE case.
Interestingly, while in the former case the emission is produced only in the
downstream, in the latter case a substantial fraction of the H$\alpha$ emission
comes from the upstream. This fact is better illustrated in
Fig.~\ref{fig:Halpha_up}, where the fraction of total H$\alpha$ emission
produced upstream is plotted as a function of shock velocity and for different
values of electron-ion equilibration efficiency upstream (while $\beta_{down}$
is fixed to 1): the upstream emission has a peak when $V_{sh}$ is close to 2500
km/s and can reach $\sim40\%$ of the total emission when $\beta_{up}= 1$.
Remarkably, even reducing $\beta_{up}$ down to 0.05, the upstream emission is
still a non-negligible fraction of the total, being around $10\%$. 
For velocities $> 2500$ km/s the upstream emission decreases monotonically because
the heating produced by the neutral return flux becomes less efficient, as
shown in Paper I (see its Fig.~6).

The H$\alpha$ emission produced upstream is clearly due to collisional
excitation of hydrogen by electrons. In fact, this process has a threshold for
$v_{rel}\simeq 2000$ km/s and peaks at $v_{rel} \simeq 3000$ km/s, hence when
the electron temperature is larger than $\simeq  1.5 \times 10^{5}$K the atomic
level $n=3$ is easily excited and H$\alpha$ emission is produced. On the other
hand, excitation produced upstream by protons is suppressed because, for a
given common temperature, protons have a thermal speed $\sqrt{m_e/m_p}$ times
lower than electrons; hence H$\alpha$ emission induced by proton
collisions and CE becomes relevant only for $T_p\gtrsim \rm few\times 10^7$K.
The heating induced by the neutral return flux may lead to such ion temperatures
upstream, but only very close to the shock, on spatial scales that are too small
to affect the H$\alpha$ emission. 

From Fig.~\ref{fig:Halpha_z} we can see that H$\alpha$ emission seems to
increase with increasing shock velocity. This is a consequence of the fact that
the production efficiency of H$\alpha$-photons is almost constant at high
$V_{sh}$. This is shown in Fig.~\ref{fig:Halpha_eff} where we plot the
efficiency of H$\alpha$ emissivity, defined as the number of H$\alpha$
photons emitted per hydrogen atom crossing the shock, namely $\epsilon=
I_{H\alpha}/(V_{sh} n_{H,0})$, where $n_{H,0}$ is the hydrogen number
density at upstream infinity. In Fig.~\ref{fig:Halpha_eff} several cases are
shown: the solid and dotted thick lines refer to NE and FE cases respectively,
while the dot-dashed thick line refers to an intermediate case (NE upstream
and partial equilibrium downstream with $\beta_{down}=0.1$). For all these
three cases we assumed that the plasma is optically thin to Ly$\beta$ emission
from hot hydrogen atoms (Case A) and optically thick to Ly$\beta$ emission from
cold hydrogen (Case B) as discussed at the beginning of \S\ref{sec:balmer}. 
In order to compare our results with previous work, we also plot the cases where
the plasma is optically thin or optically thick for both the hot and the cold
hydrogen emission, but only for the NE case.

These latter two cases can be compared with the estimated efficiency provided
by \cite{Chev-Ray78} who give $\epsilon_A= 0.048$ and $\epsilon_B= 0.27$,
independently of the shock velocity. We note that while $\epsilon_B$ is quite
close to our finding, especially for high shock speed, the value of $\epsilon_A$
is about a factor two smaller than our result. The discrepancy is probably due
to the fact that the excitation cross sections used at that time were not very
accurate.

\subsection{H$\alpha$ line profile} \label{sec:profile}
\begin{figure}
\begin{center}
\includegraphics[width=\size\linewidth]{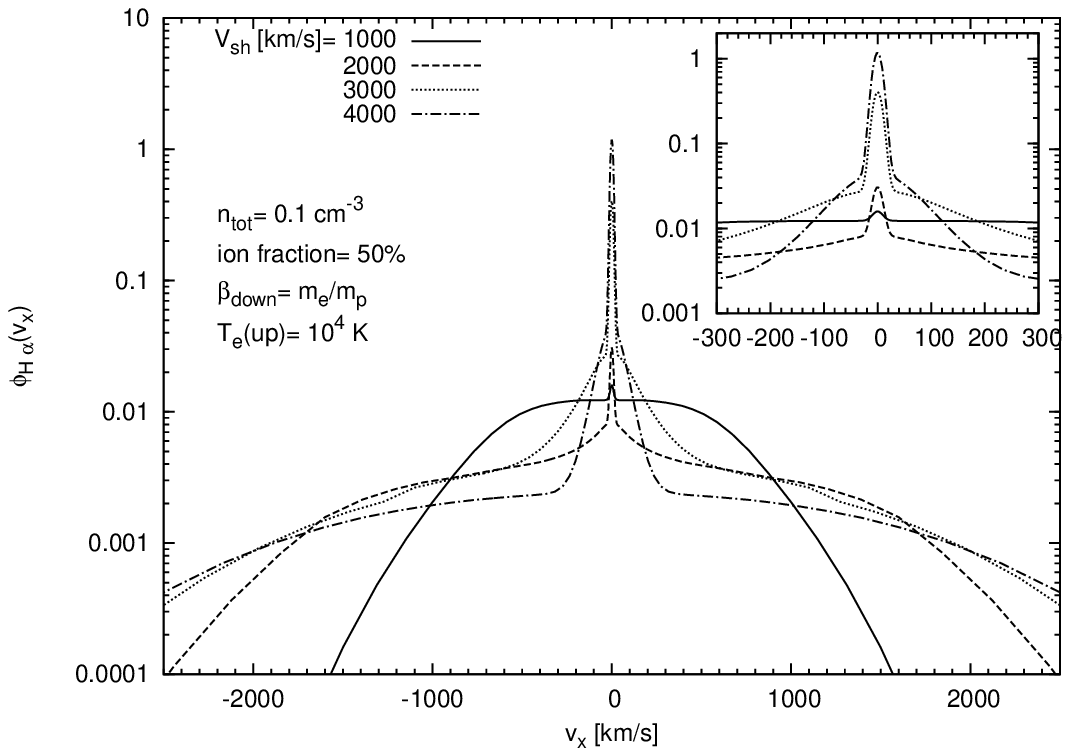}
\includegraphics[width=\size\linewidth]{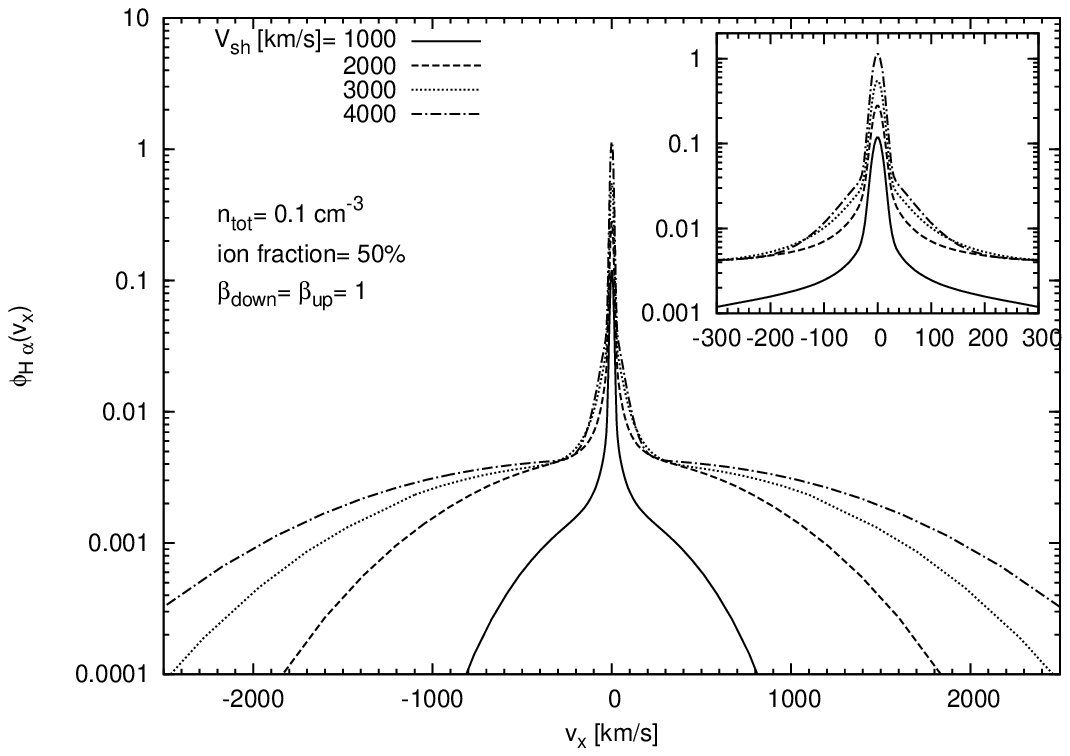} 
\end{center}
\caption{Volume-integrated H$\alpha$ emission for different values of the shock
velocity. The upper and lower panels show the cases NE and the FE cases,
respectively. The values of total density and ionization fraction are the same
as in Fig.~\ref{fig:Halpha_z}. The small boxes show a zoom for small values of
$v_x$, aimed at emphasizing the shape of the narrow and intermediate lines.}
\label{fig:Balmer-line}
\end{figure}

Let us now analyze the shape of H$\alpha$ lines. Fig.~\ref{fig:Balmer-line} shows
the volume-integrated line profiles, $\phi_{H\alpha}(v_x)$, for several values
of the shock velocity, for both NE (upper panel) and FE (lower panel) cases. 
Several comments are in order. The first point to note is that $\phi_{H\alpha}$
cannot be described, in general, by only two Gaussian-like components, as
usually assumed in the literature. Besides the usual narrow and broad
components, we clearly see the presence of a third intermediate component whose
typical width is about few hundreds km/s.
This intermediate component is a direct consequence of the existence of the
neutral-induced precursor. In fact, as we showed in Paper I, the neutral return
flux can heat upstream protons up to a temperature $\sim 10^6-10^7$K. Hence
hydrogen atoms that undergo CE upstream with these warm protons
can produce H$\alpha$ emission with a typical width of $\sim 100$ km/s.
This picture also suggests that the intermediate component should have a 
non-Gaussian profile, because it contains contributions from
hydrogen populations at different locations in the precursor, which have
different temperatures. 

\begin{figure}
\begin{center}
{\includegraphics[width=\size\linewidth]{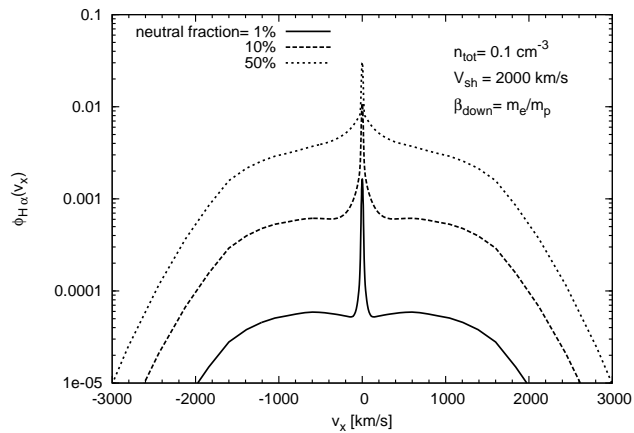}}
\caption{Volume-integrated H$\alpha$ emission for three different values of
upstream ionization fraction.}
\label{fig:Balmer-ion_frac}
\end{center}
\end{figure}

This physical interpretation of the intermediate component is supported by the
fact that its intensity, with respect to the narrow line, increases or
decreases according to the temperature and length of the neutral precursor. 
For example, in Fig.~\ref{fig:Balmer-ion_frac} we show the effect of
increasing the initial neutral fraction from 1\% to 50\%. We see that the
intermediate component becomes more prominent as the neutral fraction increases:
this is a consequence of the fact that the heating induced upstream by the neutral
return flux increases.
A similar behaviour occurs when changing the shock speed. In Paper I we showed that
the upstream temperature has a maximum for $V_{sh}\simeq 2000$ km/s and
decreases for smaller and larger speeds. The same happens for the emission of the
intermediate component with respect to the narrow one (see
Figs.~\ref{fig:fit3M1} and \ref{fig:fit3M2}).

\begin{figure}
\begin{center}
\includegraphics[width=\size\linewidth]{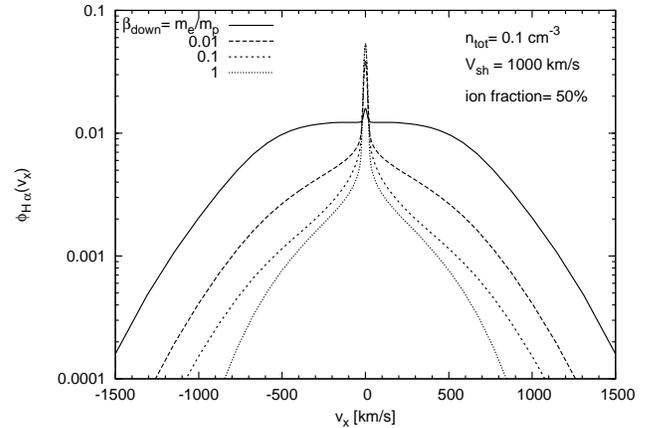}
\caption{Volume-integrated H$\alpha$ emission for different values of the
downstream electron-ion equilibration efficiency $\beta_{down}$. The upstream electron
heating is assumed to be null.}
\label{fig:Balmer1}
\end{center}
\end{figure}

\begin{figure}
\begin{center}
\includegraphics[width=\size\linewidth]{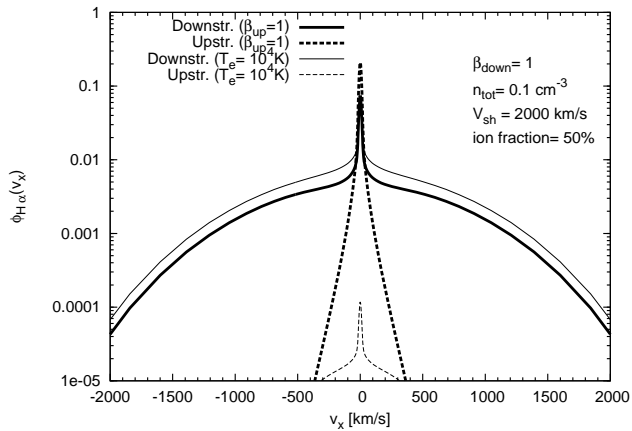}
\caption{Volume-integrated H$\alpha$ emission from downstream (solid
lines) and upstream (dashed lines) for two extreme cases of upstream electron heating:
thin lines refer to the case with constant upstream electron
temperature, $T_e=10^4$K, while thick lines are computed for complete
equilibration $\beta_{up}=1$. The value of downstream equilibration is fixed to 
$\beta_{down}=1$ for both cases.}
\label{fig:Balmer2}
\end{center}
\end{figure}

\begin{figure}
\begin{center}
{\includegraphics[width=\size\linewidth]{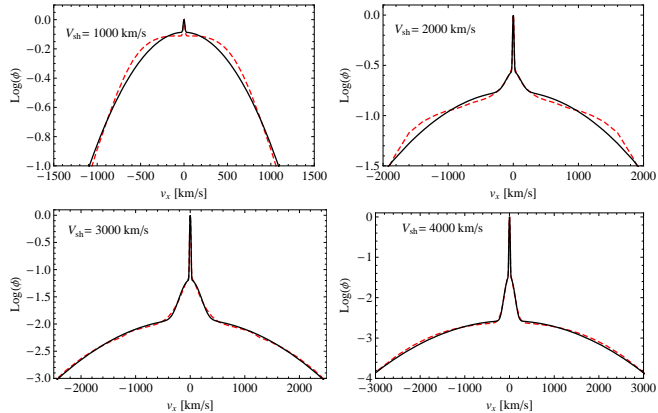}}
\caption{Examples of fits with three Gaussian curves for different values of
the shock velocity, in the NE case. The dotted lines represent the emission
computed with our kinetic model, while the solid line is the best fit. The
quality of the fit increases for larger values of $V_{sh}$. Notice that all
profiles are normalized such that the peak value is equal to 1.}
\label{fig:fit3M1}
\end{center}
\end{figure}

\begin{figure}
\begin{center}
{\includegraphics[width=\size\linewidth]{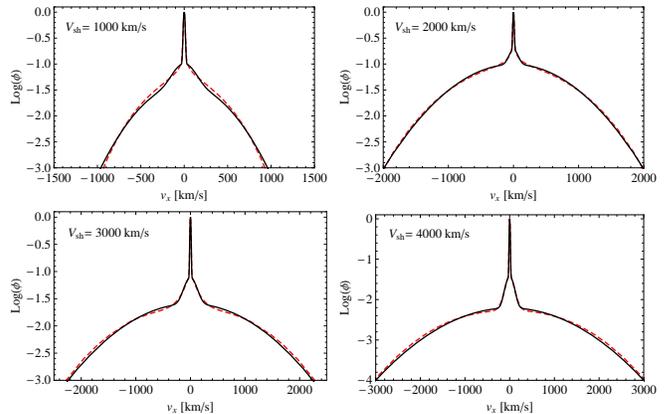}}
\caption{The same as Fig.~\ref{fig:fit3M1} but for the FE case.}
\label{fig:fit3M2}
\end{center}
\end{figure}

The variation of electron-ion equilibration efficiency also considerably affects 
the line profile. In Fig.~\ref{fig:Balmer1} we plot $\phi_{H\alpha}$ for a fixed
shock speed and for different values of $\beta_{down}$. When $\beta_{down}$
increases, a decreasing of the broad emission is observed: this occurs because
electrons contribute to ionize hot hydrogen atoms. Also the width of the broad
component decreases because a fraction of the protons' thermal energy is
transferred to electrons, hence the proton temperature decreases.
The narrow component, on the other hand, is only slightly affected by variations
of $\beta_{down}$. Its intensity increases when $\beta_{down}$ goes from
$m_e/m_p$ up to $\sim 0.01$, while for larger values it remains constant.
In Fig.~\ref{fig:Balmer1} we only consider the effect of electron-ion
equilibration downstream, while the electron temperature upstream is taken
constant and equal to $10^4$K. The effect of electron heating upstream can be
appreciated by looking at Fig.~\ref{fig:Balmer2}, where we plot separately the
volume-integrated emission from upstream and downstream assuming FE downstream,
but distinguishing the NE and the FE cases upstream. The FE case shows that the
total upstream emission is comparable to the downstream one, but has a very
different line profile, which strongly departs from a Gaussian shape.
Moreover, no broad line comes from the upstream.

In order to perform a more quantitative study of the H$\alpha$ emission with
the aim of extracting independent information from the three components, we
decide to fit the whole line profile using three Gaussian curves.
Some examples of best fit profiles are shown in Figs.~\ref{fig:fit3M1} and
\ref{fig:fit3M2} for the NE and the FE cases, respectively. The first point to
notice is that the shape of both the broad and intermediate components slightly
departs, in general, from a perfect Gaussian profile. This is especially true
for the NE case and for a shock speed value below 2500 km/s (see upper panels of
Fig.~\ref{fig:fit3M1}). On the other hand when the shock speed and/or the
electron-ion equilibration level increase, the quality of the fit improves
noticeably. We note that the deviation of the broad component from a pure
Gaussian profile was already pointed out by \cite{Heng07}.

Using the 3-Gaussian fit, we extract the FWHM of all three lines, which
provides several pieces of information. The first remarkable result is that the
width of the narrow component does not change significantly varying the shock
speed and the initial ionization fraction. Its value is always $\sim21$ km/s,
which corresponds to a population of atoms with a temperature of $10^4$ K. This
result implies that the neutral-induced precursor does not affect at all the
narrow line width, which is only determined by the hydrogen temperature at
upstream infinity. This is a consequence of the fact that the precursor length
is smaller than the CE interaction length of cold hydrogen atoms in the upstream,
irrespective of $V_{sh}$ and other ambient parameters. This result is
particularly important because it demonstrates that the neutral precursor
cannot be responsible for the anomalous narrow-line component detected from
several SNR shocks, which have a FWHM as large as $\sim$30--50 km/s.

\begin{figure}
\begin{center}
\includegraphics[width=\size\linewidth]{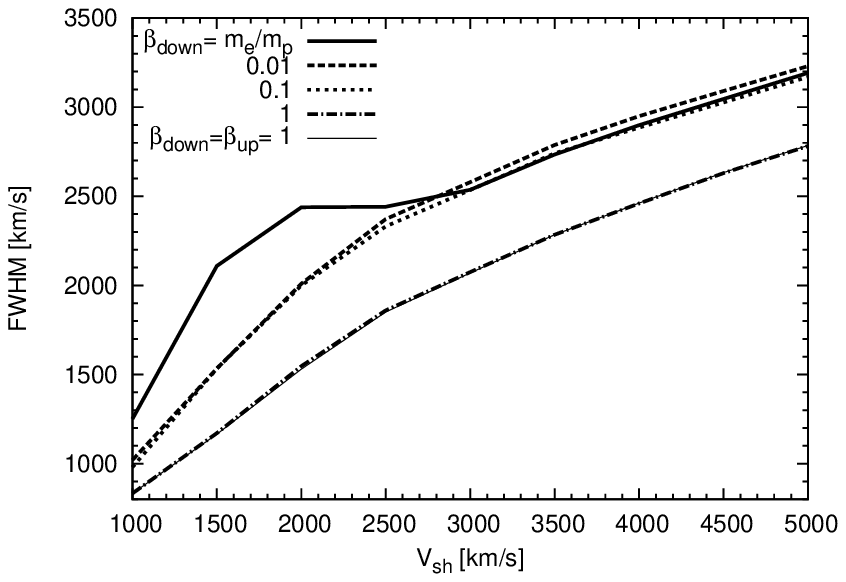}
\includegraphics[width=\size\linewidth]{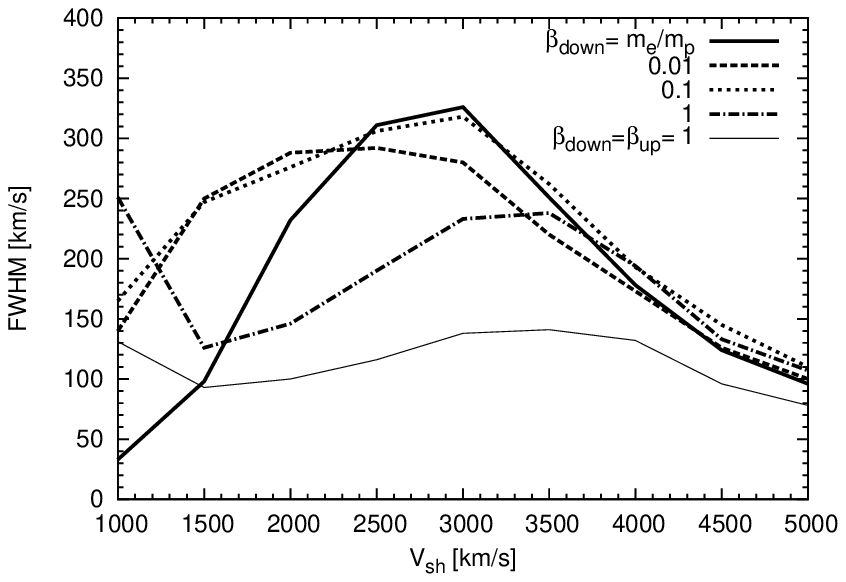}
\caption{FWHM for broad ({\it upper panel}) and intermediate components
({\it lower panel}) resulting from the the fit of total H$\alpha$ emission with
3-Gaussian distributions.}
\label{fig:FWHM}
\end{center}
\end{figure}

Concerning the broad and the intermediate components, their FWHM are shown in
the upper and lower panel of Fig.~\ref{fig:FWHM}, respectively.  As usual we
plot the results for the NE and FE cases plus some intermediate cases for the
electron-ion equilibration level. Our results for the FWHM of the broad
component are in good agreement with previous calculations
\cite[e.g.][]{Smith91}. The only exception concerns the NE case, which departs
from the general trend for $V_{sh}<2500$ km/s. This behavior does not have a
direct physical meaning and, as already noticed, is rather due to the
particularly bad quality of the fit in this region of the parameter space. We
remark that the quality of the fit rapidly improves for larger value of $\beta$
and the broad line width resulting from the fit becomes very close to the actual
width when $\beta_{down}\gtrsim 0.01$. 

As first pointed out by \cite{Chevalier80}, the FWHM of the broad line is a
direct measurement of the proton temperature downstream. As a consequence it
only depends on the values of $V_{sh}$ and $\beta_{down}$. This result can be
easily understood using a plane parallel shock model for a totally ionized plasma,
which gives a proton temperature equal to $T_i= 3 m_p V_{sh}^2 /
16(1+\beta_{down})k_B$, where $m_p$ is the proton mass and $k_B$ the Boltzmann
constant. As we showed in Paper I, this result still remains a good
approximation when the plasma is only partially ionized and the neutral return
flux is taken into account.
On the other hand, a deviation of the proton temperature from this estimate
can be induced by the presence of helium. In fact, immediately downstream of
the shock, helium nuclei thermalize at a temperature $m_{He}/m_p$ times larger
than the protons' one.
If helium and protons thermalize at the same temperature on a length scale
smaller than the excitation length scale, the mean temperature of hot hydrogen
produced by CE with hot protons is larger than the prediction without helium. As
a consequence the FWHM is expected to be larger. Indeed, the presence of helium
was taken into account by \cite{vanAdelsberg08}, which found for the broad
component a FWHM about 15-18\% larger than the one found in our calculations.

Let us now consider the width of the intermediate component (lower panel of
Fig.~\ref{fig:FWHM}). In this case, for $1000 <V_{sh}< 5000$ km/s, the FWHM
ranges between 100 and 300 km/s. A peak is present for $2000 <V_{sh}< 3500$
km/s, depending on the level of electron-ion equilibration. Once again, we notice
that for $V_{sh}<1500$ km/s the FWHM obtained from the fit procedure is not well
determined in that the intermediate component departs from a pure Gaussian
shape. Moreover, the emission due to the broad component becomes much larger than the
contribution of the intermediate one, which, in turn, becomes less
distinguishable.

Observational evidences, compatible to what we call here {\it intermediate
component}, have been reported in several works, even if such evidences have
never been related to the neutral-induced precursor. 
The most interesting case is the H$\alpha$ line profile detected from the ``Knot
g'' of the Tycho's SNR by \cite{Ghavamian00}. There, an observation of H$\alpha$
emission performed with high spectral resolution suggests the presence of two
superimposed lines: a narrow one, with a FWHM of $\sim 44$ km/s, plus a second,
less pronounced line, whose FWHM is $\sim 150$ km/s. Such a width is fully
compatible with the FWHM of the intermediate component resulting from our
calculations. On the other hand it is important to stress that the {\it
Knot g} is a complex region where density variations of the ISM produce a
non-spherical shock, hence the observed line profile could  also result from
projection effects. For this reason observations with better spatial and
spectral resolution are required in order to disentangle geometrical effects
from physical effects.
The Balmer emission detected from Tycho is not an isolated case. Several
spectra observed from Balmer-dominated shocks show evidence of narrow H$\alpha$
lines with non-Gaussian ``wings'' \cite[see e.g.][]{Smith94}. We suggest that
such wings could be the signature of the intermediate component. 

In spite of this interesting connection, in the present work we avoid performing
a detailed comparison between theoretical predictions and observations. The
reason is that although the calculations presented here represent a considerable
step forward in the description of collisionless shocks in partially ionized
media, they still remain incomplete. Aside from some minor complications that
need to be taken into account (like the presence of helium or projection effects
arising from deviations from plane geometry), a major role in determining the
shape of the H$\alpha$ line is played by the presence of CRs, which are not yet
included here. In fact, it is widely accepted that efficient CR acceleration
changes the global shock structure, generating an upstream precursor that is, to
some extent, similar to the neutral-induced precursor but develops on a
different length scale. Effects induced by CRs could be especially relevant for
the Tycho's SNR, which has been suggested to accelerate CRs efficiently
\cite[]{Mor-Cap12}. In passing we notice that the efficient acceleration of CRs
is the most plausible explanation of the relatively wide narrow  Balmer line
found in Tycho (FWHM of $\sim 44$ km/s). 

It is worth noticing that alternative explanations for the non-Gaussian wings
have also been proposed. For example, \cite{Raymond08} suggest that deviations
from the Gaussian profile could be the result of a non-Maxwellian proton
distribution downstream. In fact, neutral atoms that become ionized could
settle into a bi-spherical distribution (similar to that of pickup ions in the
solar wind) that would then introduce a non-Gaussian contribution to the line
core.
We recall that in the present work, as well as in Paper I, we do not include
such an effect, but we assume that, soon after being ionized, atoms thermalize
with the rest of ions.

\subsection{H$\alpha$ emission from the upstream} \label{sec:H_alpha_up}
This section is devoted to highlight some features of the H$\alpha$ emission
from the upstream region. This is a crucial observable in order to understand
the effects produced by CR acceleration, as we will show in a forthcoming paper.
In the near future, observations could reach good enough spatial and spectral 
resolution so as to provide detailed spectra at different distances from the
shock position, which makes the exercise of analyzing the details of the
emission from the upstream especially interesting.  

\begin{figure}
\begin{center}
\includegraphics[width=\size\linewidth]{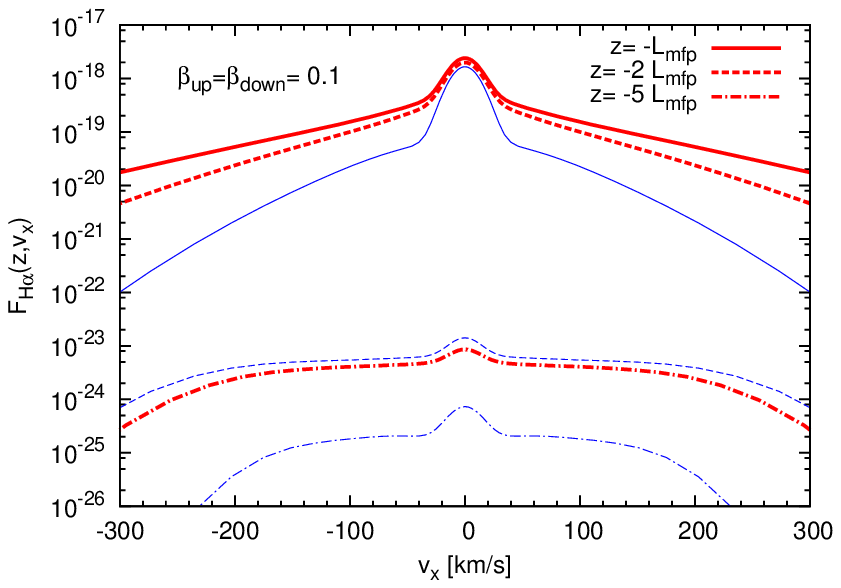}
\includegraphics[width=\size\linewidth]{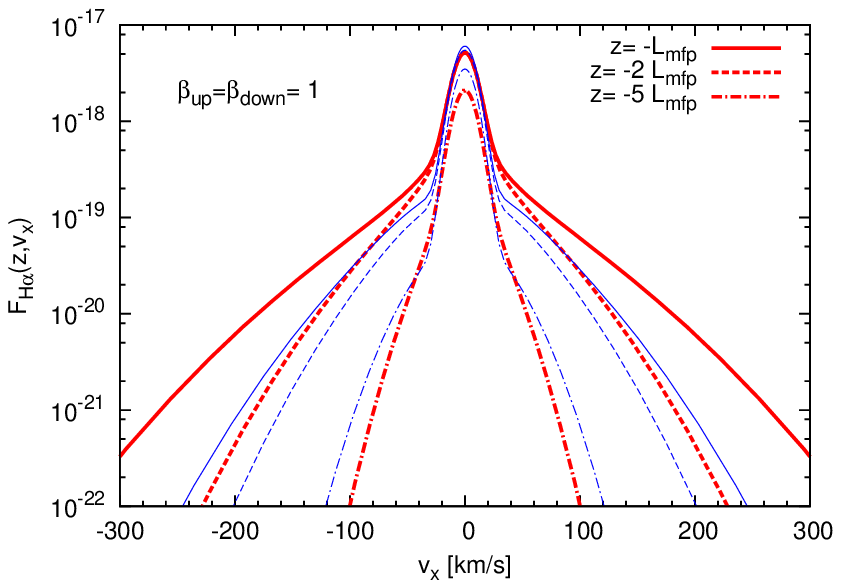}
\caption{Emissivity profile at three different locations in the upstream for
partial equilibration, $\beta_{up}=\beta_{down}=0.1$ ({\it upper panel}), and
FE, $\beta_{up}=\beta_{down}=1$ ({\it lower panel}). Thick (red) lines show the
case for $V_{sh}= 2000$ km/s while thin (blue) lines are for $V_{sh}= 4000$
km/s.}
\label{fig:Balmer3}
\end{center}
\end{figure}

\begin{figure}
\begin{center}
\includegraphics[width=\size\linewidth]{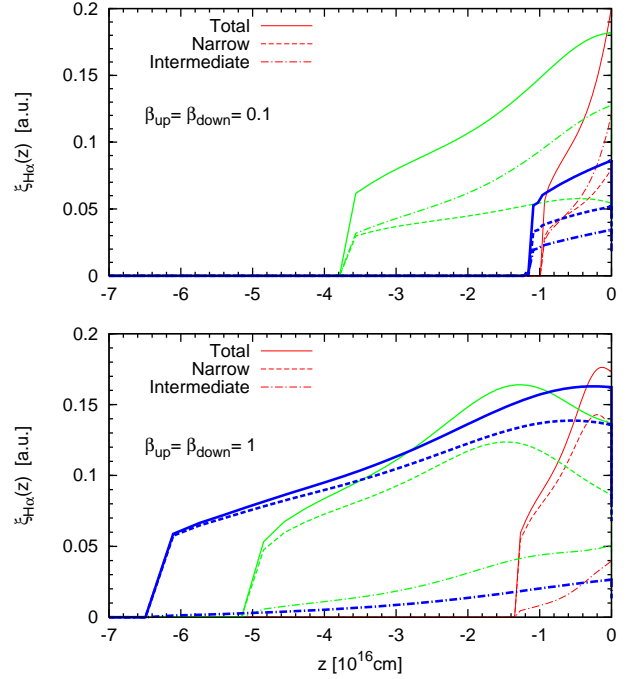}
\end{center}
\caption{Spatial emissivity profile of the H$\alpha$ line in the upstream. 
The upper panel shows the case of partial electron-ion equilibration with
$\beta_{up}=\beta_{down}=0.1$ while the lower panel shows the FE case. Dashed
and dot-dashed lines show the contribution of narrow and intermediate lines,
respectively, while the solid line is the total emission. The line thickness
(and color) distinguishes the different shock velocities: 1000 (thin-red), 2000
(middle-green) and 4000 km/s (thick-blue).}
\label{fig:Halpha_z_UP}
\end{figure}

\begin{figure}
\begin{center}
\includegraphics[width=\size\linewidth]{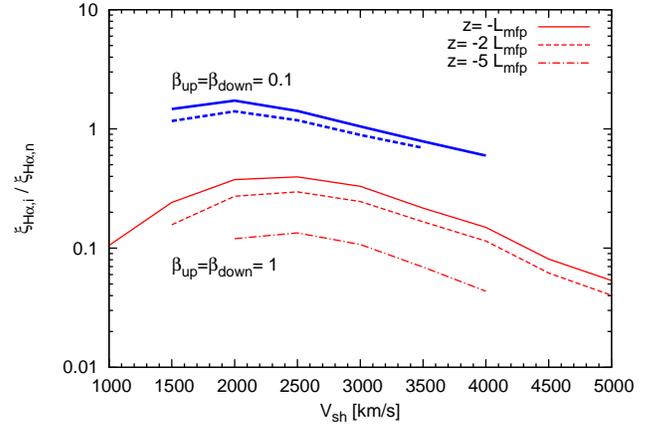}
\caption{Ratio of the intermediate to narrow H$\alpha$ emission at different
locations in the upstream, as a function of the shock velocity. Different line
shapes correspond to different locations, as indicated in the label. Thick
(blue) lines show the partial equilibration case, while thin (red) lines show
the FE case. The region of the parameter space where the H$\alpha$ emission
drops at a level of $10^{-4}$ times the peak value are not shown in the plot.
For example, for the case of partial equilibration at a distance of $5 L_{\rm mfp}$
the emission is always below this threshold.}
\label{fig:Ii-vs-In_up}
\end{center}
\end{figure}

As we already showed in Fig.~\ref{fig:Halpha_up} the upstream region could
radiate up to 40\% of the total Balmer emission in the case of full
electron-ion temperature equilibration. On the other hand if $\beta_{up}
\lesssim 0.01$ the upstream emission drops below 1\% of the total.
The line profile in the upstream emission is quite different from the one 
produced in the downstream region. In fact only the narrow and the
intermediate lines are present.
This is clearly shown in Fig.~\ref{fig:Balmer3} where the upstream line profile
at different distances from the shock is shown for the cases of FE and partial
equilibration and for $V_{sh}= 2000$ and 4000 km/s. We chose the following
distances: $z= 1\times, 2\times$ and $5\times L_{\rm mfp}$, where $L_{\rm
mfp}=1/(\sigma_{ce} n_{\rm tot})\sim 10^{16}$ cm, and the charge-exchange
cross section is approximated as $\sigma_{ce}\sim 10^{-15}$ cm$^{2}$. As we
move far away from the shock, the FWHM of the intermediate line decreases as a
consequence of a reduction of the temperature in the precursor, while the narrow
line has always the same FWHM. The total emission falls down at a distance of
$\sim$few $L_{\rm mfp}$. 
This distance corresponds to the position where the electron
temperature falls below the threshold of $\approx 1.5\times 10^5$K discussed in
the second paragraph of \S~\ref{sec:spatial}. This point moves further from the
shock for larger values of $\beta_{up}$ as can be clearly seen in
Fig.~\ref{fig:Halpha_z_UP}, where we plot the integrated line emission as a
function of the position only in the upstream, distinguishing the contribution
of the narrow and intermediate line and for different values of the shock
velocity. When we have FE the contribution of the intermediate line is always
smaller than that of the narrow line, but for lower equilibration levels the
ratio of intermediate over narrow emission increases and for $\beta_{up}= 0.1$
the two lines contribute at roughly the same level. These findings are
summarized in Fig.~\ref{fig:Ii-vs-In_up}.

\subsection{Simulated observations} \label{sec:simulations}

Although the three-Gaussian fit presented in Section \S~\ref{sec:profile}
catches the essence of the expected distorsions in the Balmer lines, it does not
reflect the whole complexity involved in fitting observed H$\alpha$ line
profiles. Actual data are affected by a number of instrumental and statistical
issues, the most obvious of which (and the only ones that we will investigate
here) are: the limited spectral resolution of the instrument, the Poisson noise
of the line photons themselves, and any additional photon noise, either due to
the astronomical or to the instrumental background.

For the simulations presented here we have used the line profile computed for
$V_{sh}=3000$~km/s (and with an ion fraction of 50\%), $\beta_{down}=m_e/m_p$,
$T_e$(up)$=10^4$~K. This situation is the one plotted as a dotted line in
the upper panel of Fig.~\ref{fig:Balmer-line}.
 
As for the observational parameters, we will adopt a published observation of
an H$\alpha$ line profile in SN~1006 \cite[]{Ghavamian02} as the reference for
the instrumental parameters as well as for the flux levels.
In that observation, the instrumental resolution is $4.5$ \AA, corresponding
to 205~km/s, while the dispersion per pixel is 0.27 times the resolution; the
total number of photons measured in the line is about $10^9$ (for a 140 min
integration time, with a 4-m class telescope), while the background noise level
is about 3000 photons/\AA: these photon numbers are for the whole spectrograph
slit and integration time, as specified by \citet{Ghavamian02}.

\begin{figure}
\begin{center}
\includegraphics[width=\size\linewidth]{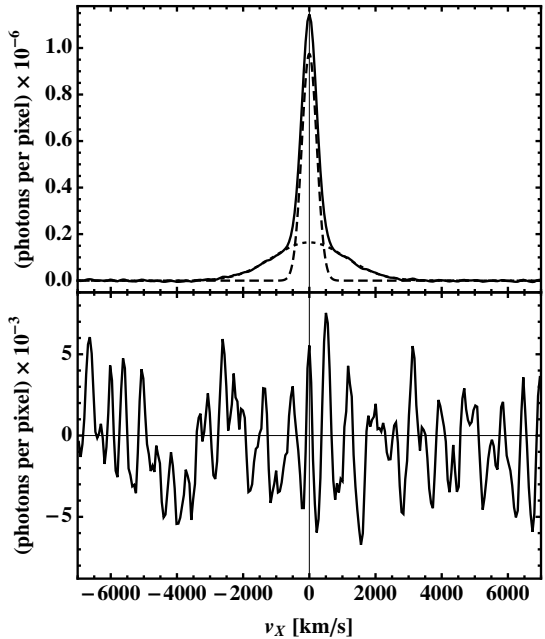}
\caption{Results of a 2-Gaussians fit to a simulated line profile, obtained by
downgrading our Model with a typical instrumental resolution and photon noise,
and adding Poisson noise (see text for details). The upper panel shows the
simulated data (solid line), together with the best-fit narrow line (dashed
line) and broad line (dotted line) components; while the lower panel shows the
residuals after subtraction of the best-fit solution.}
\label{fig:SingleSimul}
\end{center}
\end{figure}

Fig.~\ref{fig:SingleSimul} shows the results of a 2-Gaussian fit to simulated
data, obtained combining the model with the instrumental parameters mentioned
above. As shown by the residuals, in this case the quality of the observation is
not sufficient to investigate details, beyond the mere separation of a narrow
and a broad component.

We have then explored several combinations of the observational parameters,
focusing on the spectral resolution (expressed in terms of velocity resolution,
$v_{res}$) and on the photon statistics (expressed in terms of the total number
of photons in the line, $N_{phot}$).
As for the instrumental dispersion per pixel, we have kept the ratio of 0.27
times the resolution, as in \citet[][]{Ghavamian02}, while we have usually
adopted the background noise level given above. We have also tried with a much
lower background noise level, but, of course, even in this case, the noise
component associated with the photons of the line itself cannot be eliminated.

\begin{figure}
\begin{center}
\includegraphics[width=\size\linewidth]{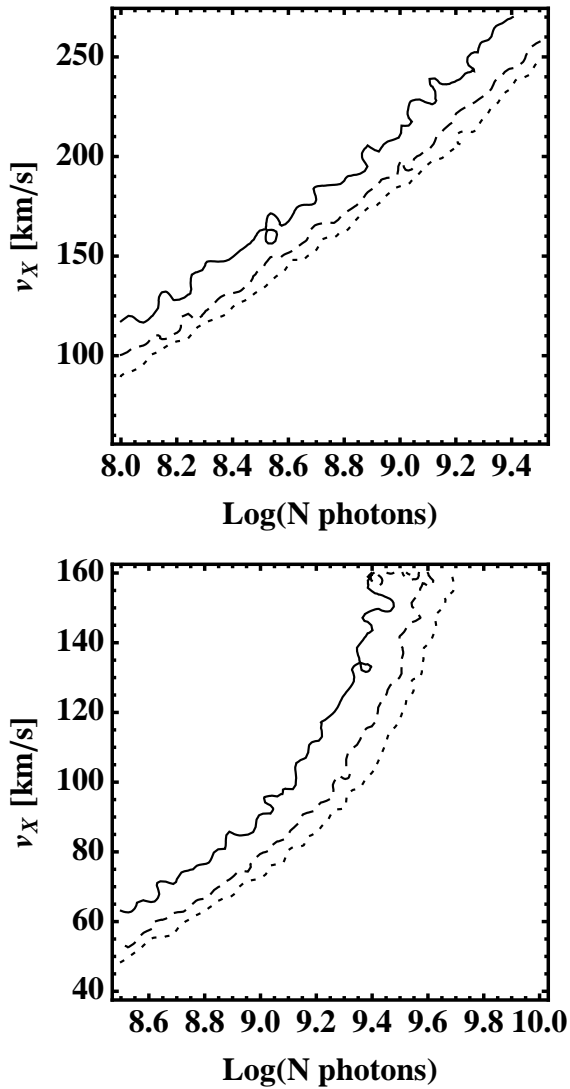}
\caption{Behaviour of the 2-Gaussians (upper panel) and  3-Gaussians (lower
panel) fits on simulated data obtained starting from our Model.  We have
investigated the parameter plane $\log_{10}(N_{phot})$--$v_{res}$. The three
contours in each panel, drawn respectively as a solid, a dashed and a dotted
line, indicate the 1-$\sigma$, 2-$\sigma$ and 3-$\sigma$ confidence level. The
contours are obtained by interpolating over a $40\times40$ data grid, each point
of which is an average value of 50 independent simulations. These plots show
that only on the lower right side of the panels the quality of the observations
is good enough to allow outlining deviations from, respectively, a 2- or
3-Gaussians profile.}
\label{fig:ParamPlane2G3G}
\end{center}
\end{figure}

We have used a grid of simulations to investigate, on the
$\log(N_{phot})$--$v_{res}$ parameter plane, the behavior of 2-Gaussian
and 3-Gaussian fits. In both cases we have chosen a grid of $25\times25$ points,
suitably positioned in the parameter plane. In order to minimize the ``noisy''
look in the figure (a natural consequence of the fact that each simulation
includes random noise), we have performed a large number (100) of simulations
for each point; out of them, we have discarded the 5 cases with the highest
$\chi^2$ value as well as the 5 ones with the lowest $\chi^2$ value, and we have
then taken the average $\chi^2$ of the remaining ones.
The results are shown in Fig.~\ref{fig:ParamPlane2G3G}.

As a result, for our Model, we may see that, for a line photon statistics of
about $10^9$ photons, a spectral resolution better than about 180~km/s is
required to show (at a 3-$\sigma$ confidence level) that a 2-Gaussian fit is
not adequate; while a resolution better than about 70~km/s is required to
challenge the 3-Gaussian fit. 
Even if having more photons does matter, in general the photon statistics does
not seem to be a parameter as crucial as the spectral resolution. Of course, in
order to resolve the narrowest spectral component a considerably higher
resolution is required; otherwise, it will be detected only as an ``unresolved
spectral component''.

\subsection{Line intensity ratios}
Another observable that may be useful in order to constrain shock parameters is
the ratio between the intensities of broad and narrow components, $I_b/I_n$.
This information is usually used in combination with the FWHM of the broad
component, in order to infer simultaneously both $V_{sh}$ and the level of
electron-ion equilibration downstream \cite[]{Heng09}. The presence of the
neutral-induced precursor complicates a bit this exercise because the upstream
equilibration also plays a role, as we show below.

At this point we need to comment on an observational caveat. When Balmer
emission is observed with a high spectral resolution, in order to resolve the
narrow component, usually the broad component is not detected due to the high
spectral dispersion \cite[se e.g.][]{Ghavamian00}. In order to measure the
intensity of the broad component, which allows one to estimate the
$I_b/I_n$ intensity ratio, observations must be performed instead with a lower
spectral resolution, typically equivalent to a velocity resolution $\Delta v
\sim 100-300$ km/s; but this does not allow to resolve simultaneously all three
components, because at such resolution the intermediate component
cannot be distinguished from the narrow one, as they appear as a single line.
As a consequence, in order to provide an estimate of $I_b/I_n$, we first 
convolve the line profile obtained by our kinetic calculation, with the typical
instrumental resolution, $\Delta v$. Then we fit the convolved emission with a
two-Gaussian profile, evaluating both $I_b$ and $I_n$ from the fitting curves.
We choose $\Delta v=150$ km/s, which corresponds to a wavelength resolution of
$\Delta \lambda= 3.3$ \AA.

\begin{figure}
\begin{center}
\includegraphics[width=\size\linewidth]{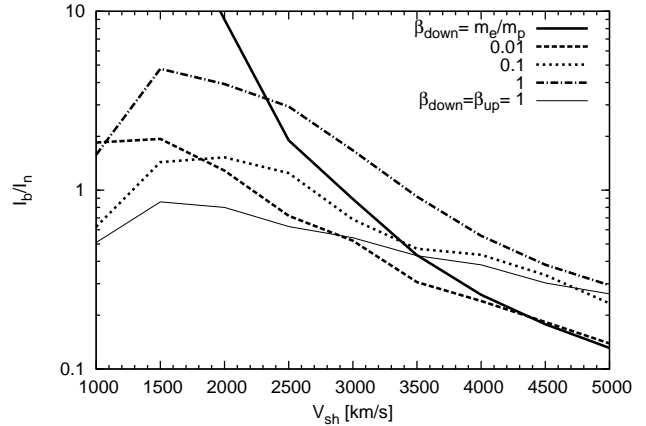}
\caption{Ratio of the broad to narrow H$\alpha$ emission versus shock velocity
after the convolution with experimental velocity resolution of $\Delta v= 150$
km/s. Different lines represent different cases of electron-ion equilibration
efficiency. Thick curves all assume non-equilibrium upstream, while the thin
curve shows the FE case.}
\label{fig:Ib-vs-In}
\end{center}
\end{figure}

The results are shown in Fig.~\ref{fig:Ib-vs-In}, where different lines
represent different assumptions for the electron-ion equilibration level.
The qualitative behaviour is similar to what predicted by previous studies
\cite[]{Smith91,Heng07}, even if some differences can be noted. Our results are,
in general, a factor 2--3 larger than what predicted by \cite{Smith91} (see
their Fig.~8). Their equilibrated case (which corresponds to our dot-dashed
line) peaks at $V_{sh}=2000$ km/s and is $I_b/I_n \simeq 2$ while our curve
peaks at $V_{sh}\simeq 1500$ km/s with $I_b/I_n \simeq 5$. 
In fact, it is rather difficult to perform a close comparison between
the ratios computed by different authors, because of substantial differences in
the model assumptions, in the methods used, in the assumed chemical composition,
and even in the cross sections adopted for the various processes: therefore we take the
above level of differences as acceptable.

From the observational point of view, in all the SNRs for which $I_b/I_n$ has
been measured, an intensity ratio above 1.2 has never been seen, while in most
cases it falls below unity \cite[see e.g.][]{Heng07}. If we assume no
equilibration upstream, this result suggests an intermediate value for the
electron-ion equilibration efficiency downstream. On the other hand we also see
that the FE model (both upstream and downstream) predicts an intensity ratio
$<1$ for all shock velocities considered. Unfortunately, for given values of
$I_b/I_n$ and $V_{sh}$ there is a degeneracy for the values of $\beta_{up}$ and
$\beta_{down}$.

Moreover, the trend of $I_b/I_n$ with respect to the electron-ion equilibration
is not monotonic. As first noticed by \cite{Smith91}, NE and FE cases do not
represent the extreme values of the intensity ratio. We see, in fact, that for
intermediate values of $\beta_{down}$, $I_b/I_n$ drops below the equilibrated
case $\beta_{down}=1$.

\section{Conclusions}
\label{sec:conc}
In this paper we computed the H$\alpha$ emission produced by a collisionless
shock that propagates into a partially ionized medium. In order to do this, we
first derived the evolution of the various species across the shock, by using
the kinetic model developed in Paper I. This model applies to plane-parallel
non-radiative shocks in the steady  state where ions are treated as a fluid,
while neutral particles are described using the full 3D velocity distribution
function. 
On top of this solution we then computed the Balmer emission produced by
collisional excitation of hydrogen atoms with both ions and electrons, as well
as by CE events leading to neutrals in excited states. Results for the spatial
emission and for the line profile of H$\alpha$ are presented for a shock seen
edge-wise, varying the shock speed, the initial ionization fraction and the
electron-ion equilibration level.

According to the traditional picture \cite[]{Chev-Ray78}, the H$\alpha$ profile 
detected from Balmer dominated shocks usually consists of two components: a
narrow one, whose width reflects the temperature of the upstream medium, and
a broad one, due to neutrals that have undergone CE with hot protons in the
downstream region.

This picture is however an oversimplification of what happens in reality, mainly
for two reasons: i) it assumes that neutrals can be described as a fluid, namely
with Maxwellian velocity distributions; ii) it does not take into account the
effect induced by the neutral precursor. In fact, the latter point is a natural
consequence of the former one: already in Paper I we showed that, for a
wide range of shock velocities, a fraction of the hot neutrals produced
downstream can recross the shock toward upstream, giving rise to a neutral
return flux, and that the interaction of these neutrals with the upstream ions
produces a precursor region where the incoming plasma is heated and slowed
down. 
In this paper we have shown that the neutral-induced precursor is responsible
for the production of a new H$\alpha$ line component, whose width is
intermediate between the narrow and the broad lines, being around few hundreds
km/s for a shock speed of a few thousands km/s. This intermediate line is due to
cold hydrogen atoms that have undergone CE with warm protons in the
neutral-induced precursor, hence its width reflects the mean temperature of the
precursor. In addition, we found that the profiles of the intermediate and
broad-line components may deviate from pure Gaussians, and that these
deviations could be detected by carrying out observations of suitably high
quality.

A natural question to ask is whether the heating produced in the neutral precursor
could explain the anomalous width of narrow lines, which has already been
observed in some Balmer filaments associated with several SNR shocks
\cite[]{Smith94, Hester94}. Our results show that this is not the case: the bulk
of incoming neutrals does not interact with ions in the neutral precursor,
because its extent, which corresponds to the interaction length of the returning
neutrals, is much smaller than the CE length of the incoming neutrals. Instead,
as we already discussed, the fraction of incoming neutrals interacting with ions
in the precursor will give rise to the intermediate H$\alpha$ line. 
Therefore, we conclude that other mechanisms, like for instance a CR-induced
precursor, should be invoked to explain an anomalous width of the narrow-line
component.

Remarkably, some observations point towards the existence of intermediate
lines compatible with our predictions: narrow lines detected from Tycho and from
other SNRs show non-Gaussian wings which could be explained with the existence
of a third line component. At the moment this result must be taken with
care because projection effects could also be responsible for the observed line
profiles. Better spatial and spectral resolution are needed to disentangle
these effects. Unfortunately, the majority of these observations do not have
the spectral resolution required to carry out a satisfactory 3-component fit. In
order to estimate the experimental requirements necessary to identify the
intermediate line, we compared our theoretical H$\alpha$ profile with
simulated observations, which take into account both instrumental resolution and
Poisson noise of the line photons. As a result, for a typical line photon
statistics of about $10^9$ photons, a spectral resolution better than $\sim70$
km/s is required to separately identify all three components. 

The presence of the intermediate line component may also affect the evaluation
of the broad to narrow line intensity ratio, $I_b/I_n$. This quantity is usually
used together with the broad line width, in order to estimate the level of
electron-ion temperature equilibration in the post-shock region. Evaluation
of $I_b/I_n$ is usually done from observations with resolution $>100$ km/s,
necessary to detect both the narrow and the broad lines.

This implies that the intermediate line is not resolved and that its emission
contributes to the observed narrow line intensity. We have included this effect
in the evaluation of $I_b/I_n$, and have shown how this ratio changes varying
the electron-ion equilibration downstream. As already pointed out by several
authors, if efficient electron-ion equilibration occurs downstream, the width of
the broad line is reduced because a fraction of the kinetic energy of incoming
protons is transferred to electrons.

In addition to the effects produced by electron-ion equilibration downstream,
we investigated what happens if equilibration also occurs in the precursor
region. Interestingly, we showed that, increasing the efficiency of
equilibration beyond a few percent, electrons can collisionally excite hydrogen
atoms, giving rise to Balmer emission also from the precursor region. For
$V_{sh}\approx$ 2500 km/s, if equilibration is complete ($T_e=T_p$), the
emission from the precursor can contribute up to $\sim40\%$ of the
total H$\alpha$
emission. This result can be instrumental to explain the results recently
published by \cite{Lee10}. They observe the Eastern limb of Tycho's SNR, finding
a gradual increase of H$\alpha$ intensity just ahead of the shock front, which
they interpret as emission from a thin shock precursor. They estimate that the
precursor emission may contribute up to 30\%-40\% of the total narrow component
emission and suggest that the precursor is likely due to CRs. In light of our
results it is clear that a correct interpretation of the pre-shock H$\alpha$
emission requires also the evaluation of the emission produced by the neutral
induced precursor.

In a forthcoming paper, currently in preparation, we will describe the theory
of collisionless shocks in partially ionized media in the presence of
accelerated particles that exert a pressure on the incoming ions. In other
words, we will generalize the non-linear theory of particle acceleration in
collisionless shocks to include the neutral return flux discussed in Paper I.
In the same paper we will calculate the shape of the Balmer lines as they are
affected by accelerated particles, and show how to use the widths of the narrow,
intermediate and broad components of the Balmer line as tools to measure
the CR acceleration efficiency in SNRs.

\section*{Aknowledgments}
We thanks an anonymous referee for his/her valuable suggestions which
allowed us to improve the quality of the present work.
We are very grateful to Kevin Heng for providing some of the cross sections used
in this work. Our work is partially funded through grants PRIN-INAF 2010 and
ASTRI.


\begin{thebibliography}{99}

\bibitem[\protect\citeauthoryear{van Adelsberg et al.}{2008}]{vanAdelsberg08}
  van Adelsberg, M., Heng, K., McCray, R., \& Raymond, J.C. 2008, ApJ, 689,
  1089

\bibitem[\protect\citeauthoryear{Balan\c{c}a et al.}{1998}]{Balanca98}
  Balan\c{c}a, C., Lin, C.~D \& Feautier, N., 1998, J Phys. B, 31, 2321

\bibitem[\protect\citeauthoryear{Barnett el al.}{1990}]{Barnett90}
  Barnett, C. F., Hunter, H. T., Fitzpatrick, M. I., Alvarez, I., Cisneros,
  C., \& Phaneuf, R. A. 1990, Collisions of H, H2, He and Li Atoms and Ions with
  Atoms and Molecules (Rep. ORNL-6086/V1; Oak Ridge: Oak Ridge Natl. Lab.)

\bibitem[\protect\citeauthoryear{Bray \& Stelbovics}{1995}]{Bray95}
  Bray, I., \& Stelbovics, A.~T. 1995, AAMP 35, 209
  
\bibitem[\protect\citeauthoryear{Belki\'c et al.}{1992}]{Belkic92}
  Belki\'c, D., Gayet, R. \& Salin, A., 1992, At. Data Nucl. Dtata Tables,
  51, 59

\bibitem[\protect\citeauthoryear{Blasi et al.}{2012}]{paperI}
  Blasi, P., Morlino, G., Bandiera, R., Amato, E., Caprioli, D., 2012, 
  ApJ, 755, 121

\bibitem[\protect\citeauthoryear{Chevalier et al.}{1980}]{Chevalier80}
  Chevalier, R.~A., Kirshner, R.~P., and Raymond, J.~C., 1980, ApJ, 235, 186

\bibitem[\protect\citeauthoryear{Chevalier \& Raymond}{1978}]{Chev-Ray78}
  Chevalier, R. A., Raymond, J. C. 1978, ApJ, 225, L27

\bibitem[\protect\citeauthoryear{Ghavamian et al.}{2007}]{Ghavamian07}
  Ghavamian, P., Laming, J.~M., Rakowski, C.~E. 2007, ApJ, 654, L69

\bibitem[\protect\citeauthoryear{Ghavamian et al.}{2000}]{Ghavamian00}
  Ghavamian, P., Raymond, J., Hartigan, P. and Blair, W.~B. 2000, ApJ, 535, 266

\bibitem[\protect\citeauthoryear{Ghavamian et al.}{2001}]{Ghavamian01}
  Ghavamian, P., Raymond, J. C., Smith, R. C., \& Hartigan, P. 2001, 
  ApJ, 547, 2005

\bibitem[\protect\citeauthoryear{Ghavamian et al.}{2002}]{Ghavamian02}
  Ghavamian, P., Winkler, P. F., Raymond, J. C., Long, K. S. 2002, ApJ, 572, 888

\bibitem[\protect\citeauthoryear{Harel et al.}{1998}]{Harel98}
  Harel, C., Jouin, H. \& Pons, B., 1998, At. Data Nucl. Dtata Tables, 68, 279

\bibitem[\protect\citeauthoryear{Helder et al.}{2009}]{Helder09}
  Helder, E. et al. 2009, Science, 325, 719

\bibitem[\protect\citeauthoryear{Heng}{2009}]{Heng09}
  Heng, K. 2009, {\it PASA}, 27, 23

\bibitem[\protect\citeauthoryear{Heng \& McCray}{2007}]{Heng07}
  Heng, K., McCray, R. 2007, MNRAS, 654, 923

\bibitem[\protect\citeauthoryear{Heng \& Sunyaev}{2008}]{Heng-Sun08}
  Heng, K., Sunyaev, R.~A. 2008, A\&A, 481, 117

\bibitem[\protect\citeauthoryear{Hester et al.}{1994}]{Hester94}
  Hester, J.~J., Raymond, J.~C. \& Blair, W.~P. 1994, ApJ 420, 721

\bibitem[\protect\citeauthoryear{Janev \& Smith}{1993}]{Janev93}
  Janev, R. K., \& Smith, J. J. 1993, Cross Sections for Collision Processes
  of Hydrogen Atoms with Electrons, Protons and Multiply Charged Ions
  (Vienna: Int. At. Energy Agency)

\bibitem[\protect\citeauthoryear{Lee et al.}{2010}]{Lee10} 
  Lee, J.~J., Raymond, J.~C., Park, S., Blair, W.~P., Ghavamian, P., Winkler,
  P.~F., \& Korreck, K. 2010, ApJ, 715, L146

\bibitem[\protect\citeauthoryear{Lim \& Raga}{1996}]{Lim96} 
  Lim, A.~J. \& Raga, A.~C. 1996, MNRAS, 280, 103

\bibitem[\protect\citeauthoryear{Morlino \& Caprioli}{2012}]{Mor-Cap12}
  Morlino, G. \& Caprioli, D. 2012, A\&A, 538, 81

\bibitem[\protect\citeauthoryear{Raymond et al.}{2008}]{Raymond08}
  Raymond, J.~C., Isenberg, P.~A. \& Laming, J.~M. 2008, ApJ, 682, 408

\bibitem[\protect\citeauthoryear{Rakowsky}{2005}]{Rakowsky05}
  Rakowsky, C.~E. 2005, AdSpR, 35, 1017

\bibitem[\protect\citeauthoryear{Smith et al.}{1994}]{Smith94}
  Smith, R.~C., Raymond, J.~C., \& Laming, J.~M. 1994, ApJ, 420, 286

\bibitem[\protect\citeauthoryear{Smith et al.}{1991}]{Smith91}
  Smith, R.~C., Kirshner, R.~P., Blair, W.~P. \& Winkler, P.~F.
  1991, ApJ, 375, 652

\bibitem[\protect\citeauthoryear{Sollerman et al.}{2003}]{Sollerman03}
  Sollerman, J., Ghavamian, P., Lundqvist, P. \& Smith, R.~C.
  2003, A\&A, 407, 249

\bibitem[\protect\citeauthoryear{Tseliakhovich et al.}{2012}]{Tseliakhovich12}
  Tseliakhovich, D., Hirata, C.~M.; Heng, K. 2012, MNRAS, 422, 2357

\end{thebibliography}
\end{document}